\documentclass[twocolumn,floatfix,groupedaddress,superscriptaddress,aps,prb,showpacs,10pt]{revtex4-1}

\usepackage{graphicx,amsfonts,amssymb,amsmath,hyperref,dsfont,color}

\begin{document}

\title{Tunable and direction-dependent group velocities in
topologically protected edge states}

\author{G\"otz S. Uhrig}
\email{goetz.uhrig@tu-dortmund.de}
\affiliation{Lehrstuhl f\"{u}r Theoretische Physik I, 
Technische Universit\"{a}t Dortmund,
 Otto-Hahn Stra\ss{}e 4, 44221 Dortmund, Germany}



\begin{abstract}
Topological effects continue to fascinate physicists since 
more than three decades. One of their main applications
are high-precision measurements of the resistivity. We propose
to make also use of the spatially separated edge states.
It is possible to realize strongly direction-dependent group velocities.
They can also be tuned over orders of magnitude so that
robust and tunable delay lines and interference devices are within reach.
\end{abstract}

\maketitle

\section{Introduction}

The edge states of topological insulators represent 
a fascinating field of research which has been founded
over thirty years ago with the discovery of the integer 
and the fractional quantum Hall effects \cite{klitz80,tsui82}.
In particular, the link to topological invariants has been
an important step forward \cite{thoul82,avron83,niu85,kohmo85}
relying in the end on the notion of gauge-invariant quantum mechanical 
phases \cite{berry84}.

Recently, the field has significantly gained interest
due to the discovery of topological insulators driven
by spin-orbit interaction, i.e., without necessity of an
external magnetic field \cite{ando13,chang13,kou14,chang15}. This experimental
progress has been anticipated in a model put forward by
Haldane \cite{halda88b} which is governed by 
non-trivial phases, but not induced by a magnetic field.
This model is still the simplest example of a tight-binding
model with non-trivial bands, i.e., with bands of non-zero
Chern number. Since then, there is an abundance of
studies of these effects, for instance the inclusion of
spin to restore time-reversal symmetry
\cite{kane05a,kane05b,hasan10}. A very recent promising progress 
is that films of solid state systems realize ferromagnetic Chern insulators
representing essentially two-dimensional lattice systems
which display the quantum anomalous Hall effect (QAHE) \cite{chang13,kou14,chang15}
\emph{without} external magnetic field. Thus, experiment has by now come
very close to the original idea of Haldane.

Our aim is to suggest two ways to put the topological
robustness of edge states to use beyond the high-precision
measurement of resistitivity in the Hall effect.
Our focus are Chern insulators which do not require 
an external magnetic field.
The idea is based on the previous finding \cite{redde16} 
that the Fermi velocity of the edge states along the edge is not a universal
quantity, but depends crucially on the details of the edge.
We aim here at a proof-of-principle illustration of the potential of this idea.
Thus we perform calculations for the archetypical model of the field,
the Haldane model on the honeycomb lattice \cite{halda88b}.
For the same reason, we do not consider spin though
the obtained results will carry over to
spin currents in Kane-Mele models \cite{kane05a,kane05b,hasan10}.

Of course, in standard quantum Hall setups the different transport behavior
of the spatially separated edge states has been discussed, for instance
in strongly differing localization \cite{feist06} or in switchable quantum 
dots coupled to single edge states \cite{feve07}. Various quantum interferometer
devices and quantum gates based on edges states have been realized \cite{karma15,yamam12}.

We put forward two effects which to our knowledge have not yet
been studied: (i) the Fermi velocity
can be made extremely direction-sensitive, i.e., it
can differ from one edge to the other by orders of magnitude.
(ii) the Fermi velocity can be tuned by suitable voltages
to vary by orders of magnitude such that one can control
transport properties realizing tunable delay lines and 
interference devices for precise measurements of delays.
Finally, we discuss various routes to realize the proposed
effects.

\section{Model}

\begin{figure}[ht]
	\centering
		\includegraphics[width=\columnwidth]{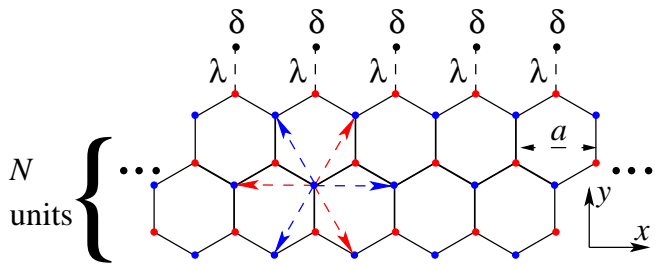}
	\caption{(Color online) Sketch of a strip of the honeycomb lattice 
	with NN hopping (black bonds) and NNN hopping (colored arrows). 
	Note that the color of the arrows is swapped 
	starting from the other sublattice.	The strip height is $N$ units. The upper
	boundary is decorated by weakly coupled ($\lambda\le1$) additional
	sites with local potential $\delta$. The lattice constant $a$
	is set to unity.}
	\label{fig:model}
\end{figure}

The tight-binding model considered reads
\begin{subequations}
\begin{eqnarray}
\label{eq:model}
H &=& H_{\rm strip}+H_{\rm decor}
\\
\label{eq:strip}
H_{\rm strip} &=& t\sum_{\langle l,j\rangle} c_l^\dag c_j
+t_2  \sum_{\langle\langle l,j\rangle\rangle} e^{i\phi_{lj}} 
c_l^\dag c_j
\\
\label{eq:decor}
H_{\rm decor} &=& \sum_{j} \left[
\lambda t(c_{d(j)}^\dag c_j + c_{j}^\dag c_{d(j)}) +
\delta c_{d(j)}^\dag c_{d(j)}
\right]
\end{eqnarray}
\end{subequations}
where the underlying lattice is shown in Fig.\ \ref{fig:model}.
The hopping on the strip of the honeycomb lattice is given
by $H_{\rm strip}$ where $\langle,\rangle$ stands
for nearest-neighbor (NN) hopping while $\langle\langle,\rangle\rangle$
stands for next-nearest-neigbor (NNN) hopping. The elements $t$ and $t_2$
are real; the former serves as energy unit. The non-trivial
topology is induced by breaking the time-reversal symmetry by
the phases $\phi_{lj}=\pm\phi$. The minus sign applies to the 
red (light gray) arrows while the plus sign applies to the blue 
(dark gray) arrows in Fig.\ \ref{fig:model}.
Starting from the other sublattice, the colors of the arrows are
 interchanged.

The Hamiltonian of the additional sites at the upper edge
is given by $H_{\rm decor}$. If $j$ is a protruding 
site at the upper edge of the honeycomb we denote its
adjacent additional site by $d(j)$. The hopping between them
is given by $\lambda t$ where $0\le\lambda\le 1$ reduces this hopping
relative to the standard NN hopping.  The local energy of the  additional sites is tuned by a gate voltage denoted by $\delta$. 
For $\lambda=0$ one retrieves the standard Haldane model on a strip of honeycomb lattice. Since this model has been studied
before \cite{halda88b,kane05b} we refrain from computing 
its non-trivial Chern numbers explicitly. We choose $t_2=0.2t$
and $\phi=\pi/2$ because this implies a sizable gap between
the Chern bands and the bands are relatively flat.

\section{Direction-sensitive Fermi velocity}

\begin{figure}[ht]
	\centering
		\includegraphics[width=\columnwidth]{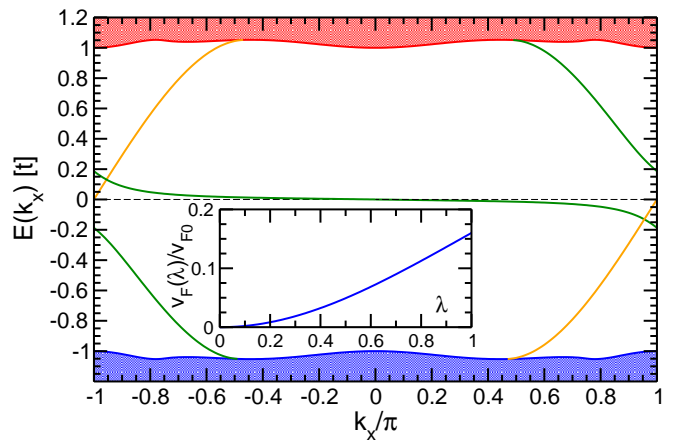}
	\caption{(Color online) Dispersion of the edge modes: in green (dark 
	gray) the left mover at the upper edge 
	and in orange (light gray) the right mover at the lower edge
	for $\lambda=0.2$.
	The shaded areas show the continua from the Chern bands.
	The inset shows the Fermi velocity $v_{\rm F}=\partial \omega/
	\partial k_x|_{\epsilon_{\rm F}}$ relative to the one at the
	undecorated edge vs.\ the relative coupling $\lambda$ at 
	$t_2=0.2t$ and $\phi=\pi/2$.}
	\label{fig:direct}
\end{figure}

We show the non-trivial properties by directly computing the
dispersion of the upper and lower edge modes. This allows
us to address their Fermi velocities. To this end,
we consider a strip of finite height, typically of $N=80$
units, see Fig.\ \ref{fig:model}. The translational invariance
in $x$-direction is preserved such that $k_x$ continues to 
be a good quantum number. Then, we diagonalize the resulting 
$(4N+3)\times(4N+3)$ matrix  numerically
for given $k_x$. This procedure yields Fig.\ \ref{fig:direct}
where we focus on the edge states. The continuum of states 
above the lower band edge $\Delta_{\rm low}(k_x):=
\min_{k_y}\omega_{\rm up}(k_x,k_y)$
are indicated by the light hatched/red colored region.
The continuum of states 
below the upper band edge $\Delta_{\rm up}(k_x):=
\max_{k_y}\omega_{\rm low}(k_x,k_y)$
are indicated by the dark hatched/blue colored region.

The left and right moving modes in Fig.\ \ref{fig:direct}
show a distinctly different behavior. This stems from the different
structure of the upper and the lower edge of the honeycomb strip
as shown in Fig.\ \ref{fig:model}. The additional decorating sites 
have a pronounced effect. Without them, the dispersion of the edge mode
crosses the Fermi level at the Brillouin zone boundary, see right
moving dispersion. If they are present, however, the Fermi level is hit at the
zone center, see left moving dispersion. 
The remarkably flat dispersion results from the weak
coupling $\lambda<1$ of the otherwise isolated decorating sites.
If $\lambda=0$ the decorating site would host a completely local,
i.e., momentum independent mode.

The dependence of the Fermi velocity $v_{\rm F}$
of the flat dispersion on $\lambda$
is studied quantitatively in the inset of Fig.\ \ref{fig:direct}.
As discussed above, it vanishes for $\lambda=0$ and grows quadratically
if the coupling is switched on. In this way, the coupling $\lambda$
provides an excellent control parameter to tune the
Fermi velocity at one edge of the sample. Since the position of the 
edge determines
the direction of motion the Fermi velocities become strongly 
direction-sensitive. In particular for small values of $\lambda$
the value for $v_{\rm F,right}$ can differ by orders of magnitude
from $v_{\rm F,left}$. This opens interesting avenues to 
applications which have not yet been realized so far. We stress that the
robustness of the topological edge mode protects them from being
destroyed or blocked {completely by disorder effects.
Of course, they will be influenced by them on a quantitative level.
But the qualitative features will persist, for instance no
localization occurs because back-scattering cannot occur.
Only if the disorder becomes very large so that scattering
from one edge to the other sets in, see for instance
Ref.\ \onlinecite{feist06}, the advocated effect vanishes.}

The direction-sensitivity of the Fermi velocity appears to
be a static property once the system is given. Next, we 
illustrate that it can be tuned as well.

\section{Tunable Fermi velocity}
The guiding idea is to continuously tune the system \emph{with}
decorating sites towards the system without them. To this end,
the decorating sites shall be switched off. This can be achieved
by pushing them up in energy so that the electrons do not visit
them anymore. The knob to do so is the local potential $\delta$
in (\ref{eq:decor}) which can be thought to be realized by a gate
voltage.

\begin{figure}[ht]
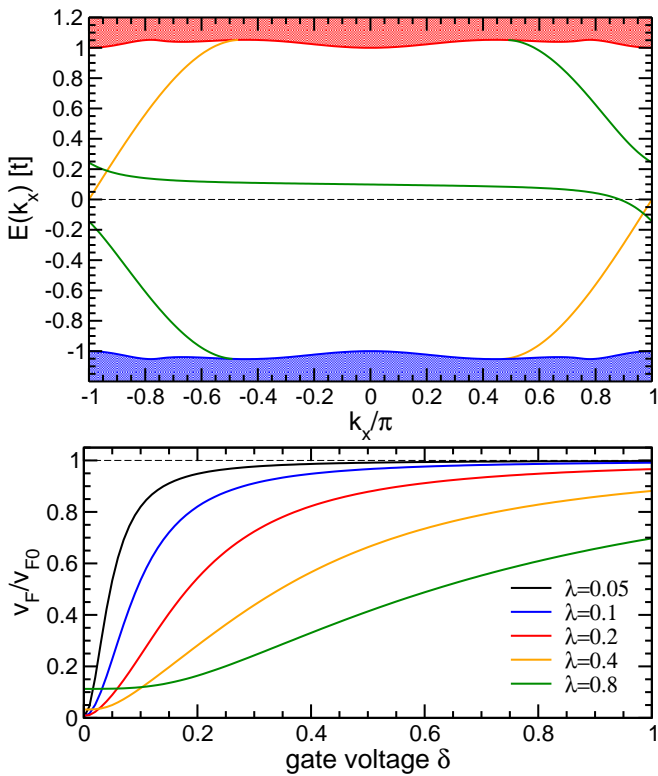

	\centering
		\includegraphics[width=\columnwidth]{fig3a}
		\includegraphics[width=\columnwidth]{fig3b}		
	\caption{(Color online) Upper panel: dispersion of the edge modes; 
	in green (dark 	gray) the left mover at the upper edge and in orange
	(light gray) the right mover at the lower edge for $\lambda=0.2$ and
	$\delta=0.1t$. The shaded areas show the continua from Chern bands.
	Lower panel: Fermi velocity 	$v_{\rm F}=\partial 	\omega/\partial k_x|_{\epsilon_{\rm F}}$ relative to the one at the undecorated edge
		vs.\ the gate voltage $\delta$ for various values of the relative coupling $\lambda$ at $t_2=0.2t$ and $\phi=\pi/2$.}
	\label{fig:tune}
\end{figure}

In the upper panel of Fig.\ \ref{fig:tune}, we depict the effect of a finite gate voltage $\delta=0.1t$. Clearly, the intended effect takes place and the
dispersion of the left moving mode is shifted upwards by about $\delta$.
In addition, the precise shape of the dispersion is modified. As expected,
the right moving mode at the other edge is almost not influenced.
The strip analysed is $N=80$ units high and for this value no effect of
the gate voltage on the right moving mode occurred, even in the nineth digit.

Due to the shift, the part of the left moving
dispersion crossing the Fermi level
at zero energy is changed. Its slope is increased as we argued
before on physical grounds. Note that due to the particle-hole symmetry
of the model a negative shift would have the same effect, i.e.,
it produces the same increase of the Fermi velocity.

In the lower panel of  Fig.\ \ref{fig:tune}, we study the increase of the 
Fermi velocity quantitatively. Indeed, it is possible
to drive the Fermi velocity to the value of the undecorated edge.
For large enough gate voltage $\delta$ the  Fermi 
velocity relative to the undecorated one saturates at unity, i.e.,
the system behaves as if the decorating sites were not present at all.
The gate voltages at which the saturation sets in is of the order
of $\lambda t$ because $\delta$ has to counteract the hybridization
between the protruding edge sites $j$ and the attached decorating
sites $d(j)$. 

This observation shows that for small values of $\lambda$ 
the gate voltage $\delta$ provides an excellent control parameter
to tune the Fermi velocity. Rather small values of the gate voltage
are sufficient to modify $v_{\rm F}$ by orders of magnitude.
There is a small offset at $\delta=0$ given by $v_{\rm F}(\lambda,\delta=0)$.
But for increasing $\delta$ a linear regime is entered in which 
$v_{\rm F}\propto \delta$ holds until the saturation regime is reached
for $\delta > \lambda t$. This linear regime is proposed to be put
to use in applications. Here the Fermi velocity, which 
is the group velocity of an electric signal, can be tuned by
a third gate so that a tunable delay line can be realized.
The gate voltage varies $\delta$ so that a setup
can be realized where the group velocity can be tuned on the fly.
In this way, the time a electric signal requires to pass through
the device can be fine-tuned to obtain destructive or constructive
interference with a signal which has passed along another path,
for instance through a sample of which the transmission properties
shall be measured. Due to the topological protection of the edge states 
disorder will not destroy the effects so that they are {qualitatively} robust
against imperfections. 

\begin{figure}[ht]
	\centering
		\includegraphics[width=0.54\columnwidth]{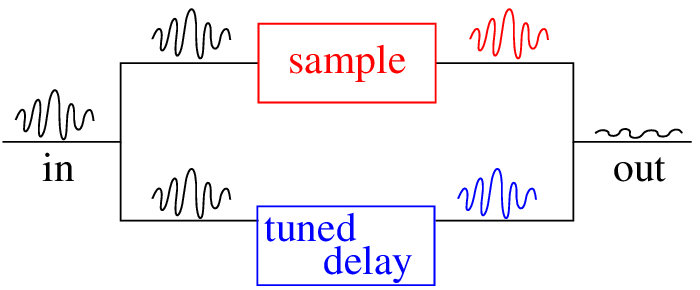}
		\quad
		\includegraphics[width=0.40\columnwidth]{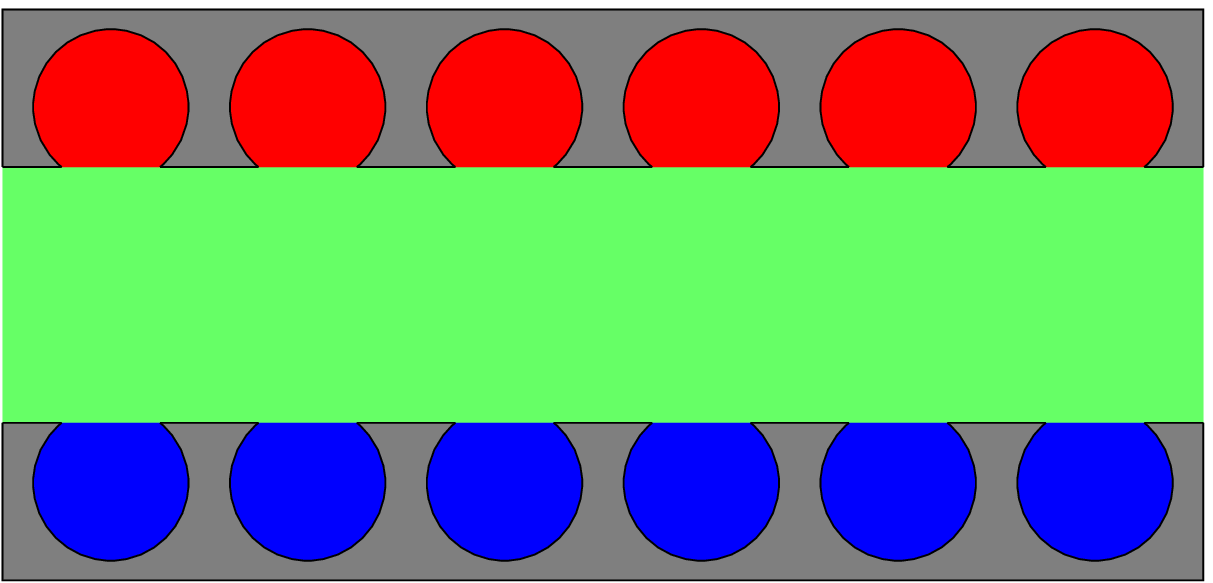}
	\caption{(Color online) Left panel: Sketch of a circuit
	to use the tunable delay line in an interference measurement
	for the determination of the delay of signal transmission
	in a sample; almost destructive interference
	is shown. Right panel: Sketch of a decorated Hall sample
	to realize tunable and direction-dependent Fermi velocities.
	The black hatched area is impenetrable for the electrons while
	the potential of the colored areas 
	(blue (dark gray) and red (light gray) can be tuned independently
	by gate voltages $\delta_{\rm low}$ and
	$\delta_{\rm up}$ while the bulk of the sample 
	(green) remains unchanged. A magnetic field is applied perpendicular
	to the plane shown to generate the edge states. 
	It should be approximately so large that
	circular Landau orbits fit into the bays.}
	\label{fig:real}
\end{figure}

In Fig.\ \ref{fig:real}a, a circuit is sketched which permits
to measure the delay occurring in the unknown sample in parallel
to the tunable delay line based on a Chern insulator with tailored
edges. The idea is to look for destructive and constructive interference
of an oscillating electric signal at the output. In this way, very precise 
detection of small delays should be possible. Note that we do
not advocate quantum interference here in contrast to many studies
in the literature, e.g., Ref.\ \onlinecite{karma15}.

Furthermore, this effect is direction-dependent. 
One can push the device one step further by decorating also the 
lower edge of the sampel (not shown). Then two independent
gate voltages can be used to control the Fermi velocities
in both directions left and right independently.

In addition, one can feed the signal itself not only to the
input of the Chern insulator, but also to the gate voltage
itself. This yields a tunable non-linear feedback: depending on the bias
on the control gate, larger amplitudes
travel faster or slower than smaller amplitudes. By suitable
tuning of this effect one can realize a wave equation similar
to the Korteweg-de Vries equation which allows for 
soliton solutions. Thus particularly stable signals can be 
transmitted.

\section{Possible realizations}

About three years ago the chances of realizing a Chern insulator
in a microscopic lattice were still considered slim \cite{bergh13}.
But the observation of the QAHE
in thin films of ferromagnetic Chern insulators has
changed the game \cite{chang13,kou14}. The temperatures at which
the QAHE is observable has increased from several milliKelvin to a 
few Kelvin \cite{chang15} and theoretical calculations indicate that
even room temperature should be within reach \cite{wu14,han15}.
Thus it appears perfectly reasonable that the proposed lattices
or similar analogues can be designed and realized. Detailed
calculations show that the tailored design of super lattices
of gold atoms on single-vacancy graphene provide a 
promising candidate to realize high temperature Chern insulators
\cite{krash11,han15}. Thus the tailored design of 
differing edges will also be possible opening a route
to realize the proposed strips of lattices and to measure
the advocated effects.

An alternative route is open in tailored optical lattices
where it has been recently
possible to realize the Haldane model (\ref{eq:model})
itself and to measure some of its basic properties \cite{jotzu14}.
Thus a proof-of-principle realization of decorated
strips of Haldane models such as shown in Fig.\ \ref{fig:model}
appears to be well within reach. While this is an attractive prospect
such a realization is probably not so close to the application as a 
measuring device.

A third alternative is to construct lattice models
artificially by design of tailored semiconductor structure.
For instance, antidot structures can already be synthesized
with high precision \cite{lan12}. So it is conceivable
to realize dot or antidot lattices which have the 
discrete translational invariance we are considering here.
Applying a perpendicular field
to such planar modulated electron gases yields the desired
topologically non-trivial Chern bands with the corresponding
gapless edge modes. We recall that lattices in a finite magnetic field
of certain strengths correspond to Haldane type of models
as illustrated before for the kagome lattice \cite{redde16}.

A fourth alternative realization can be based on more
standard quantum Hall setups. A quantum Hall sample
in a magnetic field provides a Chern insulator with spatially
separated edge states.
Thus, if one is able to pattern the edges differently and
to control them independently by gate voltages the
effect of tunable Fermi velocities will be observable.

To the author's knowledge no such experiment has been 
performed so far. But the degree of control of gate voltages
in time and space on the nanoscale is remarkable so that
it appears that the proposed experiment is well possible.
For instance, it is possible to switch the coupling of quantum dots 
to one edge mode on and off yielding electrons on demand \cite{feve07}.
If this kind of setup is extended to a periodic chain of quantum dots,
see Fig.\ \ref{fig:real}b, all gated by the same voltage, a setup is realized
which is a continuum version of the lattice model in Fig.\ \ref{fig:model}.
The patterning on the nanoscale is also possible, see for instance
the Mach-Zehnder interferometer realized by Karmakar et al.\ \cite{karma15}.
Note that the interometer discussed in that work uses the nanoscale
structures to make the two edge states interfere. Thus the setup is not
the one advocated here.

\section{Summary}

The first idea of the present Letter
is the finding that the relevant group
velocity of signals transferred in topologically
protected and thus robust edge modes may differ strongly depending on
whether they propagate at the upper or lower edge.
The reason is that the structure of the edges may differ
and that this influences the non-universal transport properties.
For disorder or a (de)coupled quantum dot such
phenomena were discussed and observed before \cite{feist06,feve07},
but not for a periodic structure with well-defined Fermi velocity
{which will allow for faithful signal transmission}.

The second idea was that this difference can be
enhanced and manipulated by special design of the edges.
For instance, weakly linked local energy levels, decorating
the edges, render the mode at the Fermi velocity arbitrarily slow.
This leads to direction-dependent group velocities because
the mode at one edge moves into one direction
while the mode at the other edge moves into the other 
direction and the link strength can be chosen very differently
at the edges.

The third idea is to tune the group velocities 
by gate voltages which can effectively switch
the weakly linked decorating levels on and off. 
By varying the gate voltages, one may easily tune the group velocities.
This can even be done for left and right movers independently.
By this mechanism the group velocity, relevant for the 
delay time in  signal transmission, can be varied by a gate voltage. 
Signals can be delayed and/or modulated.

The above ideas have been demonstrated on the proof-of-principle level
by explicit calculations
for the archetypical model for non-trival Chern bands \cite{halda88b}. 
This could be decorated by weakly coupled
additional sites at the edges with local energy
given by the gate voltage. The relative coupling $\lambda$
and the gate voltage $\delta$ are the tuning parameters.

Furthermore, we discussed and suggested ways to realize
the above proposal. Such realizations appear possible
in many fields ranging from quantum Hall systems with 
external magnetic field over ultracold atoms in optical
lattices with artificial gauges to thin films of ferromagnetic
topological insulators \cite{chang13,kou14,chang15,han15}. 
The latter represent lattice models very close to the ones
studied here theoretically.

\section{Outlook}

The findings presented have established {the fundamental idea
on the proof-of-principle level}. Depending
on the route favored for experimental realization further calculations
are called for. For instance, on the one hand 
in optical lattices the Chern insulator
is realized so far on square lattice, not on honeycomb lattice.
On the other hand, proposals in solid state physics favor honeycomb 
lattice systems which
break time-reversal symmetry by spontaneous ferromagnetism.
Thus differing specific calculations will be useful.

{The topological protection of the edges limits 
the influence of weak disorder and weak interactions 
to a quantitative level. The qualitative effect to have
tunable velocities remains robust. Nevertheless, it
would be necessary to assess quantitatively by theoretical
considerations how robust
the tunable velocities are. Such studies must be adapted to
the envisaged experimental realization; thus they cannot be made on
general level.}

The fundamental idea advocated here can be extended 
also towards spintronics. The inclusion of spin, for instance as
in the models put forward by Kane and Mele \cite{kane05a,kane05b,hasan10}
will enable to pass from currents and signals expressed in charges
to currents and signals of spin, i.e., to pass from electronics
to spintronics \cite{zutic04}.

So there appear ample ways to explore the applicability
of topologically protected edge states propagating along tailored
and tuned edges.

\acknowledgments
This work was supported by the Helmholtz Virtual Institute 
``New states of matter and their excitations'' and by TRR 160
``Coherent manipulation of interacting spin excitations in tailored semiconductors''.
Fruitful discussions are acknowledged with Manfred Bayer, Markus Betz, and
Emil J.\ Bergholtz.


%

\end{document}